\documentclass[runningheads]{llncs}
\usepackage{graphicx}
\usepackage{glossaries}
\usepackage{algorithm}
\usepackage{algorithmicx}
\usepackage[noend]{algpseudocode}
\usepackage[normalem]{ulem}
\usepackage{xcolor}
\usepackage{dsfont}
\usepackage{amsfonts}
\usepackage{bbm}
\usepackage{multirow}
\usepackage{colortbl}
\usepackage{hyperref}
\usepackage{breakurl}
\usepackage{url}
\usepackage{orcidlink}

\newcommand*\samethanks[1][\value{footnote}]{\footnotemark[#1]}

\useunder{\uline}{\ul}{}
\newacronym{dl}{DL}{Deep Learning}
\newacronym{mri}{MRI}{Magnetic Resoncance Imaging}
\newacronym{rads}{RADS}{Radiology Assessment Data System}
\newacronym{bi}{BI}{Breast Imaging}
\newacronym{fl}{FL}{Federated Learning}
\newacronym{nn}{NN}{Neural Network}
\newacronym{kd}{KD}{Knowledge Distillation}
\newacronym{ofl}{OFL}{Ordinal Federated Learning}
\newacronym{bu}{BU}{Binomial Unimodal}
\newacronym{oe}{OE}{Ordinal Encoding}
\newacronym{ce}{CE}{Cross-Entropy}
\usepackage{import}
\newacronym{cl}{CL}{Centralised Learning}
\newacronym{mae}{MAE}{Mean Absolute Error}
\newacronym{fedavg}{FedAvg}{Federated Averaging}
\newacronym{uoc}{$A_{UOC}$}{Uniform Ordinal Classification Index}
\newacronym{drl}{DRL}{Disentangled Representation Learning}
\newacronym{csd}{CSD}{Content Style Disentanglement}
\newacronym{fedgs}{FedGS}{Federated Gradient Scaling}
\newacronym{ct}{CT}{Computed Tomography}
\newacronym{hu}{HU}{Hounsfield Units}

\begin{document}
\title{FedGS: Federated Gradient Scaling for Heterogeneous Medical Image Segmentation}

\titlerunning{FedGS: Federated Gradient Scaling}
\author{Philip Schutte\inst{1}\orcidlink{0009-0005-6198-2743}\thanks{These authors contributed equally.} \and
Valentina Corbetta \inst{2,3,4}\orcidlink{0000-0002-3445-3011}\samethanks \and
Regina Beets-Tan\inst{2,4}\orcidlink{0000-0002-8533-5090}\and
Wilson Silva\inst{2,3}\orcidlink{0000-0002-4080-9328}}

\authorrunning{Schutte \& Corbetta et al.}
%
\institute{University of Amsterdam, Amsterdam, The Netherlands
\and
Department of Radiology, The Netherlands Cancer Institute, Amsterdam, The Netherlands\and
AI Technology for Life, Department of Information and Computing
Sciences, Department of Biology, Utrecht University, Utrecht, The Netherlands\\
\email{w.j.dossantossilva@uu.nl}\and 
GROW School for Oncology and Developmental Biology, Maastricht
University Medical Center, Maastricht, The Netherlands}

\maketitle              

\begin{abstract}
Federated Learning (FL) in Deep Learning (DL)-automated medical image segmentation helps preserving privacy by enabling collaborative model training without sharing patient data. However, FL faces challenges with data heterogeneity among institutions, leading to suboptimal global models. Integrating Disentangled Representation Learning (DRL) in FL can enhance robustness by separating data into distinct representations. Existing DRL methods assume heterogeneity lies solely in style features, overlooking content-based variability like lesion size and shape. We propose FedGS, a novel FL aggregation method, to improve segmentation performance on small, under-represented targets while maintaining overall efficacy. FedGS demonstrates superior performance over FedAvg, particularly for small lesions, across PolypGen and LiTS datasets. The code and pre-trained checkpoints are available at the following link: \burl{https://github.com/Trustworthy-AI-UU-NKI/Federated-Learning-Disentanglement}


\end{abstract}

\section{Introduction} \label{sec:intro}

Recently, the field of \gls{dl}-automated medical image segmentation has begun to shift towards a \gls{fl} paradigm, primarily motivated by the need to guarantee stringent privacy standards~\cite{van2014systematic,rieke2020future}. \gls{fl} offers a promising approach by enabling collaborative model training, alternating local computation and periodic communication, without sharing patient data~\cite{zhang2021survey}. 

Despite the clear advantage of not requiring data sharing, implementing FL can be challenging due to variations in data statistics among different learners. When data is uniformly distributed among participating institutions, simple methods such as iteratively aggregating clients' model parameters via a weighted average approach (i.e., \gls{fedavg}~\cite{mcmahan2017communication}) have been shown to produce effective global models, achieving performance metrics comparable to their centralised counterparts~\cite{li2019privacy}. However, heterogeneous data distributions, which naturally arise in \gls{fl} environments, pose challenges in the collaborative training process. This often results in clients overfitting to local data and underperforming on cases less represented among the clients, thus rendering parameter averaging an ineffective aggregation approach~\cite{mora2024enhancing}. 
Several works have extended \gls{fedavg} to enhance robustness against client drift (\textit{i.e.}, the state of a locally trained model drifts away from the state of the optimal global model) and address data heterogeneity. FedProx~\cite{li2020federated} introduces a proximal term to local training objectives to mitigate client drift. SCAFFOLD~\cite{karimireddy2020scaffold} uses variance reduction for this purpose. MOON~\cite{li2021model} employs contrastive learning between representations from the global model and prior local models to address client drift.

A promising alternative approach involves integrating \gls{drl} into the federated model architecture. The objective of \gls{drl} is to disentangle the underlying generative factors of the input data into distinct representations. \gls{drl} has been effectively employed in centralized settings for medical image segmentation, enhancing robustness to data heterogeneity in multi-center data~\cite{liu2022learning}. However, the integration of \gls{drl} in \gls{fl} has been explored by only a few studies~\cite{bercea2021feddis,luo2022disentangled}. Specifically, these studies implement \gls{csd}, which separates content (e.g. anatomical strutctures) from style (e.g. intensity). Then, only the content representation, assumed to be shared and consistent across centers, is employed in downstream tasks (e.g., classification, segmentation), thereby addressing data heterogeneity. 

However, \gls{csd} relies on the assumption that data heterogeneity is embedded within style features, which encode variations in acquisition protocols, scanning machines, and settings. This assumption is not universally valid, as heterogeneity among clients can also stem from the content of the images. 
Indeed, the size and shape of the target segmentation significantly influence the difficulty of the segmentation task. The variability in the complexity of target segmentations among different clients can lead to underrepresentation of challenging samples, consequently diminishing their contribution to the aggregated global model.
In lesion segmentation, lesions may present different sizes, unevenly distributed across centers (e.g., a specialized cancer institute may lack small early-stage tumors in its dataset). Smaller lesions are typically more challenging to detect, representing the cases where clinicians would benefit most from \gls{dl} segmentation models~\cite{Nair2020Exploring}. Early detection of small lesions is critical; for example, small colorectal polyps are difficult to identify but are crucial in clinical practice due to their potential for growth and malignant transformation~\cite{Ignjatovic2009Optical}. Accurate detection and management of these polyps are essential for colorectal cancer prevention~\cite{Pickhardt2013Assessment}. 

We propose a novel \gls{fl} aggregation method, \gls{fedgs}, which enhances segmentation performance on samples that are challenging due to their limited size and availability. Our key contributions are the following:

\begin{enumerate}
\item \gls{fedgs} enhances segmentation performance on small-sized, under-represented segmentation targets while maintaining overall segmentation efficacy.
\item We apply \gls{fedgs} to two segmentation models, one based on UNet~\cite{ronneberger2015u} and a state-of-the-art \gls{csd} model, called SDNet~\cite{chartsias2019disentangled}, demonstrating the effectiveness of our aggregation strategy.
\item We show improved segmentation performance compared to \gls{fedavg}, particularly for small lesions, using our proposed approach on two public lesion segmentation datasets, Polypgen~\cite{ali2023multi} and LiTS~\cite{bilic2023liver} highlighting the robustness of our method in addressing client data heterogeneity.
\end{enumerate}

\section{Methodology} \label{sec:meth}

\begin{figure}[!t] 
    \centering
    \includegraphics[width=0.8\textwidth]{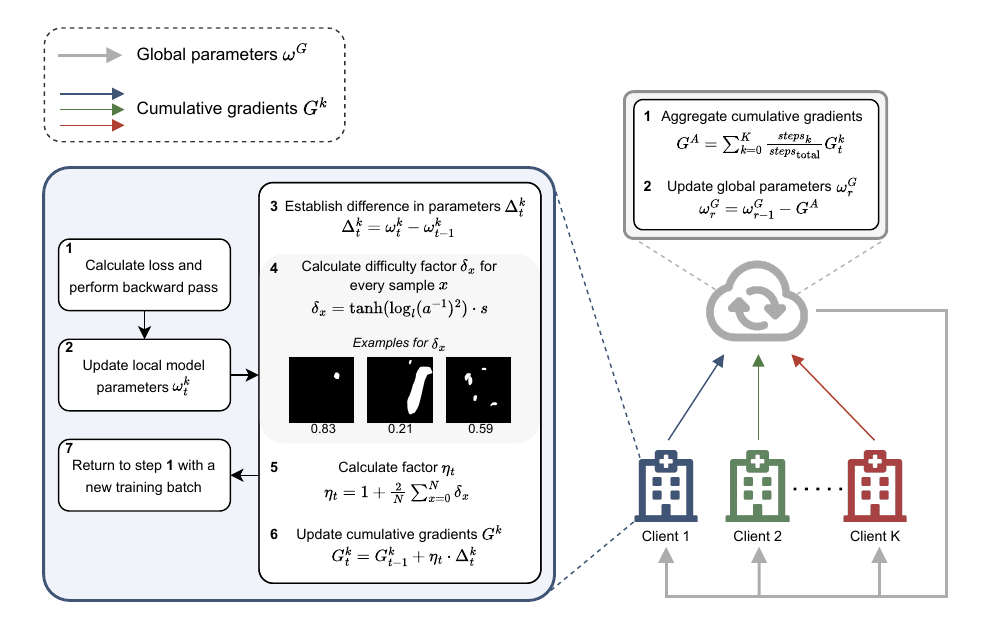}
    \caption{Overview of the FedGS aggregation method. The segmentation masks depicted in the Figure come from the PolypGen dataset.}
    \label{fig:fedgs_overview}
\end{figure}

Figure~\ref{fig:fedgs_overview} provides an overview of \gls{fedgs}. \gls{fedgs} is inspired by FedNova~\cite{wang2020tackling}, which normalizes local gradients to address inconsistencies caused by varying numbers of iterations due to differing client sample sizes. Unlike FedNova, which focuses on correcting client inconsistencies and normalizing gradients, \gls{fedgs} aims to enhance segmentation performance on challenging (i.e., small lesions) targets by scaling the gradients originating from these difficult samples. 

\subsection{FedGS overview}

In \gls{fedgs}, every client $k$ maintains a cumulative gradient $G^k_t$, where $t$ is the latest training iteration. After every client $k$ has completed its training round, the server aggregates $K$ cumulative gradients $G^k_t$ of final training iteration $T$ to obtain the aggregated cumulative gradient $G^A$. The aggregation is performed with weighted averaging based on the number of iterations completed by each client:

\begin{equation}
    \label{eq:aggr_grad}
    G^A = \sum^K_{k=0} \frac{\mathrm{steps}_k}{\mathrm{steps}_{\textrm{total}}} G^k_t
\end{equation}

The server then updates its global model parameters $\omega^G_r$ for training round $r$ using the formula $\omega^G_r = \omega^G_{r-1} - G^A$. 

During each iteration $t$ during training round $r$, client $k$ performs a standard training step which involves computing the local loss function and performing backward propagation to obtain gradient $g_t$. The local model parameters $\omega^k_t$ are then updated with learning rate $\alpha$, $\omega^k_t = \omega^k_{t-1} - \alpha \cdot g_t$. The cumulative gradient $G^k_t$ is updated with the difference in local model parameters, $\Delta^k_t = \omega^k_t - \omega^k_{t-1}$, scaled by a factor $\eta_t$, resulting in the following update rule:

\begin{equation}
    \label{eq:cum_gradient}
    G^k_t = G^k_{t-1} + \eta_t \cdot \Delta^k_t
\end{equation}

The factor $\eta_t$ is based on the estimated segmentation difficulty of the input images $X$ in the training batch. 

It is important to note that \gls{fedgs} does not rescale the gradients used to update the local model parameters $\omega^k_t$. \gls{fedgs} exclusively alters the gradients stored in the cumulative gradients, thus it does not affect the training and convergence of the local models during a training round; it only influences the aggregation of the local models at the end of a training round.

\subsection{Small lesion classification and difficulty estimation} \label{subsec:diff}

The factor $\eta_t$ is based on the estimated difficulty of the segmentation targets present in the training batch of iteration $t$. If the batch contains at least one image of a small lesion, it receives a factor $\eta_t > 1$. If no image in the batch contains a small lesion, it receives no additional scaling, thus $\eta_t = 1$. We constrain $\eta_t$ to a minimum of 1 to prevent decreasing gradients,  which would undesirably reduce the total magnitude of the accumulated gradients $G^A$ and lead to slower convergence of the global model. 

To calculate $\eta_t$, we determine a difficulty factor $\delta_x \in [0,1)$ for each image $x$ in the training batch based on corresponding mask $m^x \in \{0, 1\}^{H \times W}$ as follows:

\begin{equation}
\label{eq:diff_estimation}
\begin{gathered}
    \delta_x = \tanh(\log_{l}(a^{-1})^2) \cdot s \\
    \mathrm{where} \;\; a^{-1} = \frac{m^x_H \cdot m^x_W}{\sum_{i,j} m^x[i,j]} \;\; \mathrm{and} \;\; s=
    \begin{cases}
        1, \, \text{if } a^{-1} \geq \tau\\
        0, \, \text{otherwise}
    \end{cases}
\end{gathered} 
\end{equation}

\begin{figure}[!t] 
    \centering
    \includegraphics[width=0.8\columnwidth]{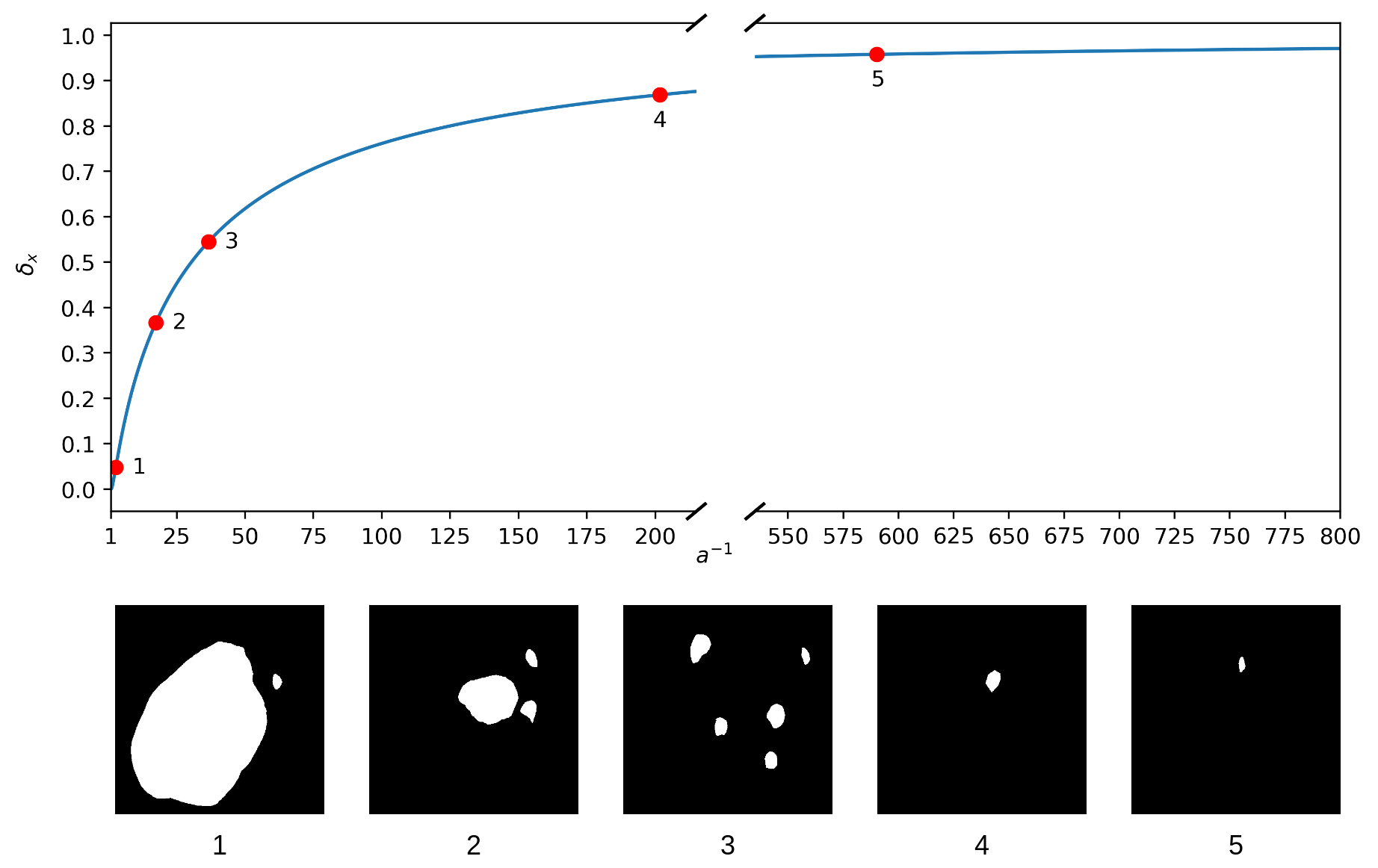}
    \caption{Plot of Equation \ref{eq:diff_estimation} that calculates the difficulty factor $\delta_x$ for images with the inverse area $a^{-1}$ as input. Additionally, five images from PolypGen that contain small polyps are shown in the graph. These images vary greatly in their total mask area and thus illustrate how the estimated difficulty changes based on $a^{-1}$.}
    \label{fig:diff_estimation}
\end{figure}

Equation~\ref{eq:diff_estimation} shows that we use the inverse relative area $a^{-1}$ of mask $m^x$ to compute $\delta_x$. This is because the inverse relative area naturally captures the overall segmentation difficulty of the lesions present in the mask. If $a^{-1}$ is equal to or greater than a predefined threshold $\tau$, then the mask contains a small lesion and $s$ is set to 1. To calculate $\delta_x$ based on $a^{-1}$, we combine the \textit{tanh} function with $\log^2_l$, transforming $a^{-1}$ to a logarithmic scale and restricting the $p_x$ values between 0 and 1. The base $l$ of $\log^2_l$ is chosen based on the overall scale of $a^{-1}$. The methods for obtaining $a^{-1}$, and the values of $\tau$ are fine-tuned for the specific segmentation task at hand. Subsection~\ref{subsec:data} details the process for the selected datasets. We have determined that $l=100$ is an appropriate setting for PolypGen, while $l=1000$ is more suitable for LiTS due to the significantly larger scale of $a^{-1}$. 
Figure~\ref{fig:diff_estimation} illustrates how $\tanh(\log_{l}(a^{-1})^2)$ responds to different magnitudes of $a^{-1}$ for $l=100$ using samples from PolypGen, aiding in visualizing the behaviour of $\delta_x$. Noticeably, the increase of $\delta_x$ rapidly decelerates as $a^{-1}$ becomes higher. This is a desirable behaviour, as it is evident from the five images that changes in mask absolute area are substantially greater at lower ranges of $a^{-1}$ than at higher ranges. There is a significant reduction in mask area from example 1 to 2, whereas the change between examples 4 and 5 is much less drastic. 

Finally, $\eta_t$ is computed by combining the difficulty factors $\delta_{x_i}$ of images $x_i$ in the training batch of size $N$ according to the following equation:

\begin{equation}
\label{eq:scaling_factor}
    \eta_t = 1 + \frac{2}{N} \sum^N_{x=0} \delta_{x_i}
\end{equation}

For images without small lesions, $\delta_x = 0$. Therefore, these images do not affect the value of $\eta_t$. Equation~\ref{eq:scaling_factor} shows that we sum all difficulty factors $\delta_{x_i}$ and scale the sum down by $\frac{N}{2}$ . We intentionally use a sum operation instead of averaging, because we want the number of small lesion samples in a batch to influence the final value of $\eta_t$. For example, a batch with three small lesion samples will have a significantly higher $\eta_t$ than a batch with only one small lesion sample, assuming that all small lesion samples have a similar $\delta_x$ value.

\section{Experiments} \label{sec:experiments}

\subsection{Model architectures} \label{ssubsec:models}

We apply \gls{fedgs} in \gls{fl} training of two \gls{dl} models: UTNet~\cite{gao2021utnet} and a modified SDNet~\cite{chartsias2019disentangled}, enhanced with a self-attention mechanism (SD-UTNet).

\subsubsection{UTNet}
Qu et al.\cite{qu2022rethinking} showed that self-attention-based architectures enhance \gls{fl} performance on heterogeneous data. We selected UTNet, a hybrid transformer architecture that integrates transformer encoders and decoders into the UNet model\cite{ronneberger2015u}.

\subsubsection{SD-UTNet}
SDNet is a state-of-the-art \gls{csd} network for medical imaging that disentangles center-invariant content from center-variant style in input images. We modified SDNet by replacing its UNet backbone with UTNet, creating the hybrid transformer \gls{csd} network, SD-UTNet.

For detailed architecture, we refer the reader to the original publications. An ablation study (Table 1, Supplementary Material) shows significant segmentation performance improvement with the introduction of self-attention.

\subsection{Datasets} \label{subsec:data}

\subsubsection{PolypGen}

PolypGen is a publicly available dataset, which contains colonoscopy data collected from 6 different centers, encompassing diverse patient populations. Our analysis focuses on single frame samples, resulting in a total of 1537 images. The datasets of centers 1-5 are used for the training and validation of the global federated models, while Center 6 is kept separate for testing. 
\paragraph{Small polyp classification} First, we need to compute the relative inverse area $a^{-1}$ of the ground truth segmentation masks $m$ to identify whether they contain at least one small polyp. To achieve this, we apply a blob detector~\cite{kong2013generalized} on $m$ to locate the center points of the polyps. The blob detector includes a visual module that identifies regions in the image that stand out from the background. In cases where a small polyp is attached to a larger polyp, the blob detector alone is insufficient. Thus, we perform an erosion operation on the mask before applying the blob detector to separate attached polyps. Finally, for the center point with the lowest estimated scale, we use its estimated size to compute $a^{-1}$. If the inverse area exceeds the threshold $\tau$, the blob is classified as small polyp. For PolypGen, we set $\tau = 150$, based on the dataset authors' description~\cite{ali2023multi}. This classification reveals that in all 6 centers, small samples are a minority class, as shown in the histogram in Figure 1(a) of the Supplementary Material.

\subsubsection{LiTS}

LiTS~\cite{bilic2023liver} is a liver tumor segmentation dataset with 131 \gls{ct} scans. We perform the experiments in 2D by extracting slices with non-empty tumor masks from the 3D volumes and saving them as 2D images, clipping \gls{hu} values to the range $[-200, 200]$. We divided the scans into five centers: centers 1-4 each with 27 scans (24 for training, 3 for validation), and center 5 with 23 scans for testing.
\paragraph{Small tumor classification} In classifying the small tumors, we notice that liver tumors cover a significantly smaller image area on average than the polyps from PolypGen. Consequently, the scale of $a^{-1}$ is higher than for PolypGen. Therefore, we choose base $l=1000$. Moreover, the threshold is set $\tau=1000$, with small tumors comprising around 20\% of all samples, as can be seen in the histogram in Figure 1(b) of the Supplementary Material. 
Finally, we exclude the blob detection step used in small polyp classification. Instead, we compare $a^{-1}$ of the entire mask against $\tau$. This approach prevents misclassification caused by small blobs that appear as small tumors in 2D slices but are actually part of a larger tumor when viewed in 3D across adjacent slices. By excluding blob detection, we prevent these masks from being incorrectly classified as "small."

\subsubsection{Pre-processing} 

All images are resized to $512 \times 512$ pixels. PolypGen images are normalized using ImageNet mean and standard deviation, while LiTS images are normalized with a mean of 0 and standard deviation of 1. Augmentations applied with a 0.5 probability include horizontal and vertical flips, and random rotations up to 90 degrees. Additionally, for PolypGen, color jitter is applied with a 0.3 probability.

\begin{table}[!t]
\caption{FedGS segmentation performance compared against FedAvg for SD-UTNet and UTNet on PolypGen and LiTS. The highest score per metric for each model is highlighted in \textbf{bold}. Equivalent scores are \underline{underlined}. Results for PolypGen are reported with the standard deviation across the 5 folds.}
\label{tab:res}
\begin{tabular}{lllccclcccll}
\cline{1-10}
                             &                                                                       &  & \multicolumn{3}{c}{PolypGen}                                                                                                         &  & \multicolumn{3}{c}{LiTS}                                                                                  &  &  \\ \cline{4-6} \cline{8-10}
\multirow{-2}{*}{Model}      & \multirow{-2}{*}{\begin{tabular}[c]{@{}l@{}}FL\\ Method\end{tabular}} &  & DiceS                                      & DiceL                                      & Dice                                       &  & DiceS                                   & DiceL                          & Dice                           &  &  \\ \cline{1-10}
                             & FedAvg                                                                &  & 0.43±0.04                                  & \underline{0.77±0.01}                                  & 0.72±0.01                                  &  & 0.4266                                  & \textbf{0.6336}                & \textbf{0.6041}                &  &  \\
\multirow{-2}{*}{SD-UTNet} & \cellcolor[HTML]{EFEFEF}FedGS                                         &  & \cellcolor[HTML]{EFEFEF}\textbf{0.44±0.03} & \cellcolor[HTML]{EFEFEF}\underline{0.77±0.01} & \cellcolor[HTML]{EFEFEF}\textbf{0.73±0.01} &  & \cellcolor[HTML]{EFEFEF}\textbf{0.4806} & \cellcolor[HTML]{EFEFEF}0.6189 & \cellcolor[HTML]{EFEFEF}0.5991 &  &  \\
                             &                                                                       &  & \multicolumn{1}{l}{}                       & \multicolumn{1}{l}{}                       & \multicolumn{1}{l}{}                       &  & \multicolumn{1}{l}{}                    & \multicolumn{1}{l}{}           & \multicolumn{1}{l}{}           &  &  \\
                             & FedAvg                                                                &  & 0.39±0.02                                  & \textbf{0.77±0.01}                         & \textbf{0.72±0.02}                                  &  & 0.4287                                  & \textbf{0.6561}                & \textbf{0.6237}                &  &  \\
\multirow{-2}{*}{UTNet}    & \cellcolor[HTML]{EFEFEF}FedGS                                         &  & \cellcolor[HTML]{EFEFEF}\textbf{0.41±0.03} & \cellcolor[HTML]{EFEFEF}0.76±0.01          & \cellcolor[HTML]{EFEFEF}0.71±0.01          &  & \cellcolor[HTML]{EFEFEF}\textbf{0.4499} & \cellcolor[HTML]{EFEFEF}0.6390 & \cellcolor[HTML]{EFEFEF}0.6120 &  & 
\end{tabular}
\end{table}

\subsection{Training setup} \label{subsec:train}
Both models are trained for 500 epochs (100 rounds of 5 epochs of local training) for PolypGen and 300 epochs (60 rounds of 5 epochs of local training) for LiTS. For PolypGen, we perform 5-fold cross-validation, saving the best checkpoint for each fold. We then evaluate each checkpoint on the test set and report the average performance of the five models. Due to time and computational constraints, 5-fold cross-validation was not feasible for LiTS; however, the large size of the testing set provides sufficiently representative results. We use a batch size of 4 for both datasets and the AdamW optimizer with a learning rate $\alpha$ of 0.0001. SD-UTNet is trained using the original implementation's loss functions, while UTNet is trained with Dice loss. We train both models using our aggregation strategy, \gls{fedgs}, and also \gls{fedavg} for comparative analysis.

\subsection{Results} \label{subsec:res}

Model performance is assessed by the Dice Score. This metric is divided into three components: total Dice Score (Dice), DiceS (computed only on lesions classified as small), and DiceL (computed on the remaining lesions). Empty masks are excluded from the computation of DiceS and DiceL. Results are reported in Table~\ref{tab:res}. We observe that \gls{fedgs} generally enhances performance for smaller lesions (DiceS) across both datasets. For larger lesions (DiceL) and overall Dice, \gls{fedgs} exhibits performance that is consistent with or marginally better or worse than \gls{fedavg}. The improvement in DiceS is more pronounced on the LiTS dataset.  
This discrepancy is likely because samples in PolypGen containing a large polyp and at least one small polyp are classified as small, as illustrated in Figure~\ref{fig:diff_estimation}, whereas in LiTS, all samples classified as small exclusively contain small lesions. Consequently, the stable performance on large polyps may account for the smaller improvement observed. Figure 2 in the Supplementary Material corroborates our findings qualitatively. Table 2 of the Supplementary Material shows that FedGS introduces a marginal training runtime overhead of 5.6\% to 13.4\% compared to FedAvg, justified by significant improvements in segmentation performance, especially for small and under-represented lesions.

\section{Conclusion} \label{sec:conclusion}

We have introduced a novel aggregation strategy, \gls{fedgs}, designed to address heterogeneity arising from varying and under-represented sizes of the segmentation targets. \gls{fedgs} has demonstrated its effectiveness in enhancing segmentation performance for small polyps and tumors while maintaining overall segmentation quality. Our results indicate that the gradient scaling approach of \gls{fedgs} is particularly effective for datasets with high variability in mask size, such as PolypGen, as well as for datasets characterized by significantly smaller mask sizes, such as LiTS. Compared to other \gls{fl} aggregation strategies, such as those introduced in Section~\ref{sec:intro}, or techniques like oversampling and class weighting for minority classes during training, \gls{fedgs} offers the advantage of leaving local training processes unaffected, thereby simplifying implementation. This approach reduces complexity by avoiding the need for additional modifications or hyperparameter tuning during local training. It maintains scalability, as the standard local training process can be uniformly applied across all clients regardless of their number. Additionally, it enhances global model performance by focusing improvements on the aggregation strategy, addressing challenges like class imbalance and client drift without altering local training dynamics. Although our evaluation of \gls{fedgs} has been confined to medical image segmentation, we believe that \gls{fedgs} holds potential for successful application in other imaging tasks and even beyond the imaging domain. This potential is contingent upon the ability to estimate the difficulty of the data with respect to the downstream task.

\subsubsection*{Acknowledgements}
\footnotesize
Research at the Netherlands Cancer Institute is supported by grants from the Dutch Cancer Society and the Dutch Ministry of Health, Welfare and Sport. The authors would like to acknowledge the Research High Performance Computing (RHPC) facility of the Netherlands Cancer Institute (NKI).

%
%
\bibliographystyle{splncs04}
\bibliography{mybibliography}

\begin{thebibliography}{10}
\providecommand{\url}[1]{\texttt{#1}}
\providecommand{\urlprefix}{URL }
\providecommand{\doi}[1]{https://doi.org/#1}

\bibitem{ali2023multi}
Ali, S., Jha, D., Ghatwary, N., Realdon, S., Cannizzaro, R., Salem, O.E., Lamarque, D., Daul, C., Riegler, M.A., Anonsen, K.V., et~al.: A multi-centre polyp detection and segmentation dataset for generalisability assessment. Scientific Data  \textbf{10}(1), ~75 (2023)

\bibitem{bercea2021feddis}
Bercea, C.I., Wiestler, B., Rueckert, D., Albarqouni, S.: Feddis: Disentangled federated learning for unsupervised brain pathology segmentation. arXiv preprint arXiv:2103.03705  (2021)

\bibitem{bilic2023liver}
Bilic, P., Christ, P., Li, H.B., Vorontsov, E., Ben-Cohen, A., Kaissis, G., Szeskin, A., Jacobs, C., Mamani, G.E.H., Chartrand, G., et~al.: The liver tumor segmentation benchmark (lits). Medical Image Analysis  \textbf{84},  102680 (2023)

\bibitem{chartsias2019disentangled}
Chartsias, A., Joyce, T., Papanastasiou, G., Semple, S., Williams, M., Newby, D.E., Dharmakumar, R., Tsaftaris, S.A.: Disentangled representation learning in cardiac image analysis. Medical image analysis  \textbf{58},  101535 (2019)

\bibitem{gao2021utnet}
Gao, Y., Zhou, M., Metaxas, D.N.: Utnet: a hybrid transformer architecture for medical image segmentation. In: Medical Image Computing and Computer Assisted Intervention--MICCAI 2021: 24th International Conference, Strasbourg, France, September 27--October 1, 2021, Proceedings, Part III 24. pp. 61--71. Springer (2021)

\bibitem{Ignjatovic2009Optical}
Ignjatovic, A., East, J., Suzuki, N., Vance, M., Guenther, T., Saunders, B.: Optical diagnosis of small colorectal polyps at routine colonoscopy (detect inspect characterise resect and discard; discard trial): a prospective cohort study. The Lancet. Oncology  \textbf{10 12},  1171--8 (2009). \doi{10.1016/S1470-2045(09)70329-8}

\bibitem{karimireddy2020scaffold}
Karimireddy, S.P., Kale, S., Mohri, M., Reddi, S., Stich, S., Suresh, A.T.: Scaffold: Stochastic controlled averaging for federated learning. In: International conference on machine learning. pp. 5132--5143. PMLR (2020)

\bibitem{kong2013generalized}
Kong, H., Akakin, H.C., Sarma, S.E.: A generalized laplacian of gaussian filter for blob detection and its applications. IEEE transactions on cybernetics  \textbf{43}(6),  1719--1733 (2013)

\bibitem{li2021model}
Li, Q., He, B., Song, D.: Model-contrastive federated learning. In: Proceedings of the IEEE/CVF conference on computer vision and pattern recognition. pp. 10713--10722 (2021)

\bibitem{li2020federated}
Li, T., Sahu, A.K., Zaheer, M., Sanjabi, M., Talwalkar, A., Smith, V.: Federated optimization in heterogeneous networks. Proceedings of Machine learning and systems  \textbf{2},  429--450 (2020)

\bibitem{li2019privacy}
Li, W., Milletar{\`\i}, F., Xu, D., Rieke, N., Hancox, J., Zhu, W., Baust, M., Cheng, Y., Ourselin, S., Cardoso, M.J., et~al.: Privacy-preserving federated brain tumour segmentation. In: Machine Learning in Medical Imaging: 10th International Workshop, MLMI 2019, Held in Conjunction with MICCAI 2019, Shenzhen, China, October 13, 2019, Proceedings 10. pp. 133--141. Springer (2019)

\bibitem{liu2022learning}
Liu, X., Sanchez, P., Thermos, S., O’Neil, A.Q., Tsaftaris, S.A.: Learning disentangled representations in the imaging domain. Medical Image Analysis  \textbf{80},  102516 (2022)

\bibitem{luo2022disentangled}
Luo, Z., Wang, Y., Wang, Z., Sun, Z., Tan, T.: Disentangled federated learning for tackling attributes skew via invariant aggregation and diversity transferring. arXiv preprint arXiv:2206.06818  (2022)

\bibitem{mcmahan2017communication}
McMahan, B., Moore, E., Ramage, D., Hampson, S., y~Arcas, B.A.: Communication-efficient learning of deep networks from decentralized data. In: Artificial intelligence and statistics. pp. 1273--1282. PMLR (2017)

\bibitem{mora2024enhancing}
Mora, A., Bujari, A., Bellavista, P.: Enhancing generalization in federated learning with heterogeneous data: A comparative literature review. Future Generation Computer Systems  (2024)

\bibitem{Nair2020Exploring}
Nair, T., Precup, D., Arnold, D.L., Arbel, T.: Exploring uncertainty measures in deep networks for multiple sclerosis lesion detection and segmentation. Medical image analysis  (2020). \doi{10.1016/j.media.2019.101557}

\bibitem{Pickhardt2013Assessment}
Pickhardt, P., Kim, D.H., Pooler, B., Hinshaw, J., Barlow, D.S., Jensen, D.W., Reichelderfer, M., Cash, B.: Assessment of volumetric growth rates of small colorectal polyps with ct colonography: a longitudinal study of natural history. The Lancet. Oncology  \textbf{14 8},  711--20 (2013). \doi{10.1016/S1470-2045(13)70216-X}

\bibitem{qu2022rethinking}
Qu, L., Zhou, Y., Liang, P.P., Xia, Y., Wang, F., Adeli, E., Fei-Fei, L., Rubin, D.: Rethinking architecture design for tackling data heterogeneity in federated learning. In: Proceedings of the IEEE/CVF conference on computer vision and pattern recognition. pp. 10061--10071 (2022)

\bibitem{rieke2020future}
Rieke, N., Hancox, J., Li, W., Milletari, F., Roth, H.R., Albarqouni, S., Bakas, S., Galtier, M.N., Landman, B.A., Maier-Hein, K., et~al.: The future of digital health with federated learning. NPJ digital medicine  \textbf{3}(1), ~119 (2020)

\bibitem{ronneberger2015u}
Ronneberger, O., Fischer, P., Brox, T.: U-net: Convolutional networks for biomedical image segmentation. In: Medical image computing and computer-assisted intervention--MICCAI 2015: 18th international conference, Munich, Germany, October 5-9, 2015, proceedings, part III 18. pp. 234--241. Springer (2015)

\bibitem{van2014systematic}
Van~Panhuis, W.G., Paul, P., Emerson, C., Grefenstette, J., Wilder, R., Herbst, A.J., Heymann, D., Burke, D.S.: A systematic review of barriers to data sharing in public health. BMC public health  \textbf{14}(1), ~1--9 (2014)

\bibitem{wang2020tackling}
Wang, J., Liu, Q., Liang, H., Joshi, G., Poor, H.V.: Tackling the objective inconsistency problem in heterogeneous federated optimization. Advances in neural information processing systems  \textbf{33},  7611--7623 (2020)

\bibitem{zhang2021survey}
Zhang, C., Xie, Y., Bai, H., Yu, B., Li, W., Gao, Y.: A survey on federated learning. Knowledge-Based Systems  \textbf{216},  106775 (2021)

\end{thebibliography}

\end{document}